\documentclass[10pt,aps,epsf,prb,preprint]{article}
\usepackage{graphicx}
\usepackage{psfig}
\usepackage{epsfig}
\begin{document}

\title{Gravitational Lensing In Case of Massive Particles}
\author{Nikhil Jayant Joshi$^a$\footnote{Electronic Address: njoshi@tifr.res.in}\\{\small $^a$Tata Institute of Fundamental Research, Mumbai, India - 400 005}}

\date{}
\maketitle

\begin{abstract}
Gravitation lensing calculations, which are generally done for light ray, are extended to that for a massive particle. Many interesting results were observed. We discuss the scattering cross section along-with many consequential quantities here. In particular, the case of Schrwarzschild metric was taken as illustration, though the analysis applies to a wide range of cases, such as extended black holes.  

\end{abstract}

\section{Introduction}
Recently, there has been a lot of interest toward fixing the mass of a neutrino\cite{Kraus:ku}\cite{Pascoli:2003ke}, which constitute a large part of the cosmic background. It is anticipated, that such a neutrino flux can be used to probe the gravitational potential of the scattering object. In conventional relativity text books, a calculation of such a process is done with the case of light ray (particle with vanishing rest mass), which is the simplest possible, though very important case of gravitational scattering. Even though, for massive neutrinos moving with large momentum, light-like approximation should work with no much deviations from the actual experimental results, the case we discuss is also of importance, since it is observed that there is no proffered range of energies with which cosmic background particles come in and hence, one should not {\it a priori} be driven by the assumption of getting high energetic particles always. 
\par
In the present paper, we discuss the various quantities relevant in experimental observations of such a scattering process for a massive particle. In particular we define two quantities, namely magnification ratio ($\mu$) and refractive index ($\eta$). Following standard notations, magnification ratio, in general, can be defined as the ratio of the apparent area, projected on the source plan, of the measured flux to the area of the incoming flux on the same plane. In this work, we try to attach a refractive index to the space, following similar ideas from the usual optics. Refractive index, is defined as the proper instantaneous velocity of the particle at a spatial point to its asymptotic value. It can be anticipated, that since defining above quantities use of only geometric structure of the system is made, the quantities are applicable as soon as metric, and hence geometry of the space-time is given. It is not possible to give a general formula for these quantities, as can be seen from their metric dependence and/or choice of the co-ordinates. We illustrate the idea by deriving these quantities for the specific metric we have chosen, Schwarzschild metric.
\par
Many interesting, new features were observed, compared to the case of light scattering. The analysis turns out to be of no much complication, but instructive, in this sense. The main course, of the paper, is divided into various sections. In section {\bf II} we start with deriving scattering results for the case of Schwarzschild metric. In section {\bf III}, we discuss the magnification ratio for this case. Refractive index for this metric is studied in section {\bf IV}.

\section{Scattering Calculations}
The geodesic equation for the present case\footnote{here, we mean Schwarzschild metric} (static, isotropic, single compact center) is given by\cite{wei},

\begin{equation}
A(r)\left(\frac{dr}{dp}\right)^2 +\frac{J^2}{r^2} - \frac{1}{B(r)} = -E
\end{equation}
The constants of motion (i.e. $E$ \& $J$), bear the same physical meaning as in the classical regime, i.e. $J$ is angular momentum and $(1-E)/2$ is energy in the classical regime. 
\par
The analysis will go on the same line as for the case of light (as per the standard textbook calculations), but this time we have more complications. Unlike the case for light, we are bound to keep both $J$ and $E$ non-zero. However, as mentioned in the introduction, this doesn't make the analysis very compli     cated and one can easily proceed further.
\par
Eliminating the parameter '$p$' from the above equations and substituting {\bf Robertson Expressions} for $A(r)$ and $B(r)$ (both truncated up to second terms, following standard analysis\cite{wei}), we end up with a relation between the '$\phi$' co-ordinate and '$r$' co-ordinate as
\begin{equation}
\phi(r) = \int_{r_{min}}^r \frac{\left(1 + \gamma\frac{GM}{r}\right)\left(\frac{J}{r^2}\right)dr}{\sqrt{(1 - E) + \frac{2GM}{r}-\frac{J^2}{r^2}}}
\end{equation}
One could take higher approximations in $A(r)$ without much difficulty. $B(r)$ seems to be more troublesome, but it can be seen that, there is no harm in including one higher order, which would change our definition of $J$, in further analysis. But, since one expect a very small correction even at the higher energies(and even at the speed of light!), there is no need to worry about the corrections introduced by higher approximations.\\
The above integral splits into nice parts, one equal to the classical expression and the next as a relativistic correction,
\begin{equation}
\phi(r) = \int_{r_{min}}^r \frac{\left(\frac{J}{r^2}\right)dr}{\sqrt{(1 - E) + \frac{2GM}{r}-\frac{J^2}{r^2}}} + \gamma GM \int_{r_{min}}^r \frac{\left[\frac{J}{r^3}\right)dr}{\sqrt{(1 - E) + \frac{2GM}{r}-\frac{J^2}{r^2}}}
\end{equation}
Small algebraic rearrangements give, in the limit $r\to \infty$
\begin{equation}
\phi_0 = \left(1+\gamma\frac{G^2M^2}{J}\right)\int_{r_{min}}^r \frac{\left(\frac{J}{r^2}\right)dr}{\sqrt{(1 - E) + \frac{2GM}{r} - \frac{J^2}{r^2}}} + \gamma\frac{GM}{4}\sqrt{1-E}
\end{equation}
The integral in the first term is just the classical expression\cite{lan}. The multiplicative factor contains the relativistic correction, which is of the order of ($G^2M^2/J$),  and which decreases as impact parameter increases. This is expected, since far away from the source\footnote{property due to compactness of the source}, space-time is essentially flat. On the other hand, this correction blows up at $J=0$ limit. It can be seen from the comparisons shown in fig.2, that this quantity cease to be meaningful immediately after it reaches approximately the limiting value set by the Schwarzschild radius. One can not interpret the result by giving any simple meaning to the results then, since all the equation and hence the analysis is invalid in the region given by $r \leq R_{Sch}$. The comparison between bending of light ray and a massive particle shows that there is a clear cut distinction between the two cases (fig.3)

\begin{figure}[h!]
\begin{center}
\epsfig{file=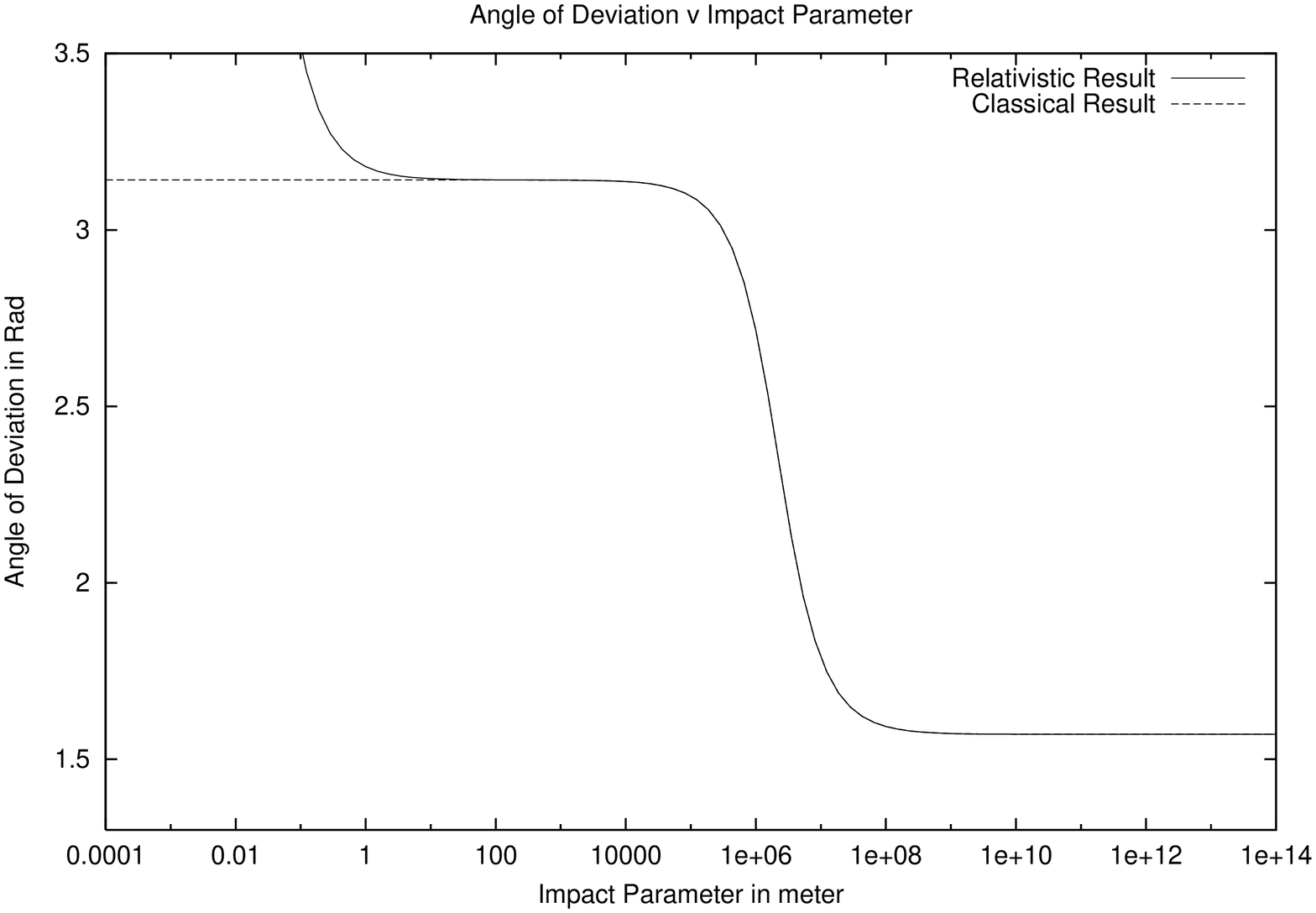,angle=-90,scale=0.25}\epsfig{file=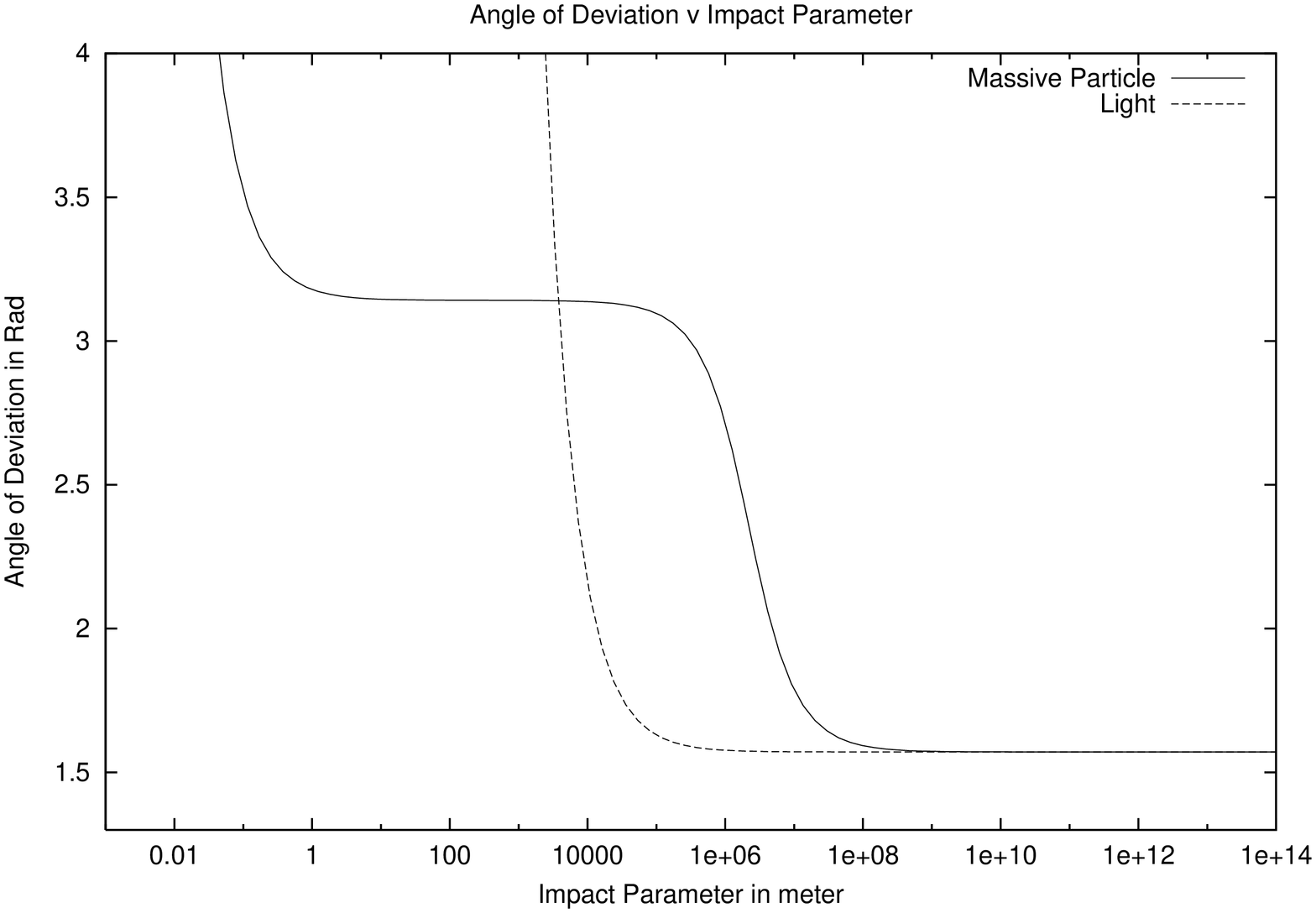,angle=-90,scale=0.25}
\caption{fig a. Comparison between classical and relativistic results for a massive particle around Sun-like center, fig b. Comparison between Angle of bending for light and that of a massive particle}
\end{center}
\end{figure}

Having calculated $\phi_0$, it is straightforward to get $\chi$, and hence differential cross section $d\sigma$, since the relation\footnote{since, the problem reduces to an equivalent central potential problem} between differential cross section\cite{lan} and $\chi$, is same as that for Coulomb scattering.i.e.
\begin{equation}
d\sigma = k'^2 \frac{d\Omega}{\sin^4(\chi/2)}
\end{equation}
where, $k' = GM/(1-E)$ and $d\Omega$ is the solid angle.
\par
On the other hand, it should be noted that, because\footnote{the situation can be thought as the effect of an attractive potential} of the possibility of multiple images, in general the experimental measurements are more involved. Following figure illustrates this point.
\begin{figure}[h!]
\begin{center}
\includegraphics[angle=0,scale=0.3]{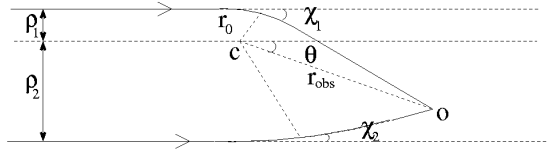}
\caption{Various components in the flux due to Multiple Images}
\end{center}
\end{figure}

 The no. of particles arriving the observer per unit time per unit area of cross section as O is given by,
\begin{equation}
dn = \sum_i dn_i = \sum_i \frac{\rho_i}{r_o\sin\theta}\left|\frac{d\rho}{d\theta}\right| d\Omega
\end{equation}
Summation is over each value of $\rho$, satisfying
\begin{equation}
\rho_i + r_o\sin\theta = \left(r_o\cos\theta - \rho_i \tan(\chi_i/2)\right)\tan \chi_i
\end{equation}
since $\chi$ is a known function of $\rho$, the above equation can be solved for allowed values of $\rho$ for given $\theta$ and $r_{obs}$. {\it (Note that the exact expression for the flux will contain the components of $dn_i$'s in the direction $\theta$, but in the limit $r_{obs} \to \infty$ the corrections are insignificant)}
\\
In the similar fashion one can get the relation between $r$ and $t$ as,
\begin{equation}
t(r,r_{min}) = \int^r_{r_{min}} \frac{[1+(2+\gamma)\frac{GM}{r}] dr}{\sqrt{\frac{1}{B(r)}-E-\frac{J^2}{r^2}}}
\end{equation}
which again can be separated into two parts, classical and relativistic correction as,
\begin{eqnarray}
t(r,r_{min}) &=& (\left[\frac{\sqrt{(r^2-r^2_{min})-2a(r-r_{min})}}{\sqrt{(1-E)}}\right] \\
&&+ (1+\gamma)\frac{GM}{(1-E)^{3/2}}ln\frac{(r-a)+\sqrt{(r^2-r^2_{min})-2a(r-r_{min})}}{r_{min}-a})\nonumber 
\end{eqnarray}
with now $a=(GM)/(1-E)$. Compared to the classical result\footnote{Classically, one gets
\begin{eqnarray}
t(r,r_{min}) &=& \frac{\sqrt{(r^2-r^2_{min})-2a(r-r_{min})}}{v_{\infty}} \nonumber\\
&&- \sqrt{\frac{ma^3}{k}}ln\frac{(r-a)+\sqrt{(r^2-r^2_{min})-2a(r-r_{min})}}{r_{min}-a}\nonumber 
\end{eqnarray}
which is nothing but
\begin{eqnarray}
\tau = t(r_1,r_{min}) + t(r_2,r_{min}) - \frac{r_1 + r_2}{v_\infty} \nonumber
\end{eqnarray}

} there is a clear-cut demarcation, now. Classically, the particle takes less time to travel between any two points (symmetric to $r_{min}$) in presence of the scattering center than it would have taken if it were moving on the straight line joining the two points when center is now there, whereas relativistically, the particle takes more time. This result is quite interesting, since it can be utilized as a check for validity of the relativistic domain. In the presence of the field, the particle traveled more distance between two spatial points, but with a velocity (of magnitude) always greater than $v_\infty$ i.e. it traveled greater path with a greater velocity. In classical case, the additional path introduced due to bending must be such that the particle took less time, whereas in relativistic case, there are two differences. The path length in this case is more (because of Robertson approximation) and the particle has to travel more distance than the classical one\footnote{Unlike classical case, now the work done on the particle by the field is used to 1. increase its speed and 2. increase its inertia}. At the same time there is an effect of time dilation, which was absent in the classical case.

\section{Magnification Ratio}
The magnification ratio is defined as the ratio of the projection of the apparent area of the flux of the particles in presence of the scattering center to that if it were not there. Since, we assume the incoming particle flux to be of uniform areal density, it can be shown that this quantity equals, the ratio between the apparent area of the flux to area through which the flux was incident. The idea is illustrated in fig.2
\begin{figure}[h!]
\begin{center}
\includegraphics[scale=0.6,angle=0]{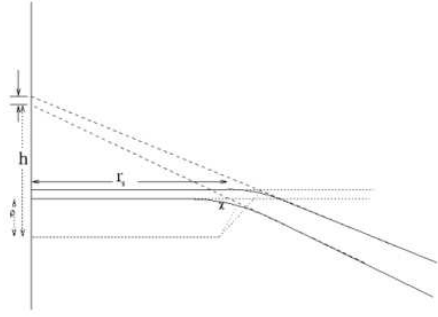}
\caption{Apparent path diagram (analogous to ray diagram) for the particle in a cylindrically symmetric system.}
\end{center}
\end{figure}

In the case (see fig.2), when there is cylindrical symmetry (which in most cases can be assumed without trouble), it is given by
\begin{equation}
\mu = \frac{h}{\rho}\frac{dh}{d\rho}
\end{equation}
 Further, from the geometry of the figure ($\chi = 2\phi_0 - \pi$), we have
\begin{equation}
h = \rho + r_s \tan \chi(\rho)
\end{equation}
Given metric, $\chi$ can be expressed in terms of $\rho$, the differential $\frac{dh}{d\rho}$ and hence $\mu$, is well defined (since $\chi$ is a monotonically varying function of $\rho$).
\par
In our case we get,
\begin{equation}
\mu = 1-\frac{4(k'/\rho)^3(r_s/\rho)}{(1-(k'/\rho)^2)}-\frac{4(k'/\rho)^2(r_s/\rho)^2}{(1-(k'/\rho)^2)^2}
\end{equation}
where, $k'= (GM/v_\infty^2)$. The relation between differential cross section and magnification ratio can be given by,
\begin{equation}
d\sigma = \frac{2\pi}{\mu}\left[\frac{2k'}{(1-\tan^2(\chi/2))} + r_s\tan^2(\chi)\right]dr_s
\end{equation}
The above result is expected, since the differential cross section is a function of boundary conditions and possibly of metric only, whereas $\mu$ does depend upon $r_s$, the distance between object plane from the scattering center.

\section{Refractive Index} 
We define local refractive index of a spatial point as the ratio of the magnitude of the asymptotic velocity of the particle ($v_{\infty}$), to the magnitude of the velocity at that point. In the classical case, where the above definition is unambiguous, the relativistic case needs more specification, as there are more than one velocities. It can be shown that it is easier to work with the proper velocity of the particle. Hence,

\begin{equation}
\eta (x,y,z) = \frac{v_{\infty}}{v_{proper}(x,y,z)}
\end{equation}
From the symmetry of the Schwarzschild metric, it can be guessed that the refractive index will not depend upon $\phi$ and $\theta$ co-ordinates. The main concern is, whether it is possible to determine the trajectory of the particle completely, once the initial conditions are supplied. Because, only then one will be able to talk about the equivalence between specifying the metric and specifying the refractive index.
\par
We start with the classical result,
\begin{equation}
\phi= \int_{r_{min}}^r \frac{(J/r^2)dr}{\sqrt{2m[E-\frac{k}{r}]-\frac{J^2}{r^2}}}
\end{equation}
which under proper b.c. leads to the expression,
\begin{equation}
\cos\phi=\frac{J/r +mk/J}{\sqrt{2mE+m^2k^2/J^2}}
\end{equation}
Hence,
\begin{equation}
\sin\phi=\frac{\sqrt{2m[E-k/r]-\frac{J^2}{r^2}}}{\sqrt{2mE+\frac{m^2k^2}{J^2}}}
\end{equation}
Now, if we define $\eta(r)$ as,
\begin{equation}
\eta(r) = \frac{\sin(\phi\to\infty)}{\sin(\phi(r))}
\end{equation}
which is nothing but the typical {\bf Snell's law} in optics, then from eq.14, it can be seen that this definition matches with our definition in terms of the local velocity.
\par
For relativistic case, one would go along the same line and it is again easy to verify that our definition is equivalent to that obtained by Snell's relation. And from our experience, in optics, that once initial angle of incidence is given for light ray, using Snell's law one can determine the path completely, one can deduce that, once impact parameter of the particle is given, which is equivalent to specifying the angle of incidence, one can use the refractive index to determine the trajectory. In case of particle, one more information must be provided, namely incident velocity (which is equivalent to specifying momentum and hence in case of light ray, specifying its wavelength, which will determine the opacity of the material).
\par
In this case, one can define the refractive index as
\begin{equation}
\eta = v_\infty/v = \frac{\sqrt{(1-E)}}{\frac{E}{A^{1/2}(r)}\sqrt{1/B(r) - J^2/r^2-E}}
\end{equation}

\section{Conclusions}
From the above treatment, it can be deduced, that, that neutrinos, if found to be massive, will follow a trajectory different from that of light ray will, corrections depending upon the value of the impact parameter. The time delay measurement done accurately enough, can be used to test the validity of the energy domain for application of classical results. On the other hand, the later analysis shows that, the analogy between gravitational lenses and optical lenses can be extended further to include refraction effects.

\section{acknowledgments}
I would like hereby express my humble gratitude toward Prof. D. Narasimha for offering me an opportunity to study this problem. I would also be thankful to Ambar Jain for his valuable comments in this work.

\end{document}